\documentclass[10pt,journal]{IEEEtran}
\usepackage{amsmath, amssymb, amsopn}
\usepackage{mathrsfs}
\usepackage{graphicx}
\usepackage{psfrag}

\usepackage{amssymb}
\usepackage{epsfig}
\usepackage{research5}

\newtheorem{lemma}{Lemma}

\newtheorem{theorem}{Theorem}

\newtheorem{definition}{Definition}

\begin{document}
\newcommand{\Markov}{{-\hspace{-.635em}\circ\hspace{.4em}}}
\newcommand{\bml}[1]{\mbox{\boldmath $ #1 $}}
\newcommand{\be}{\begin{equation}}
\newcommand{\ee}{\end{equation}}
\newcommand{\bea}{\begin{eqnarray}}
\newcommand{\eea}{\end{eqnarray}}
\newcommand{\beaa}{\begin{eqnarray*}}
\newcommand{\eeaa}{\end{eqnarray*}}

\thispagestyle{empty}
\title{The State-Dependent Multiple-Access Channel with States Available at a Cribbing Encoder }

\author{Shraga~I.~Bross, and Amos Lapidoth
\thanks{S. Bross is with the School of Engineering, Bar-Ilan University,
 Ramat Gan, 52900, Israel.
 Email:brosss@macs.biu.ac.il.}\thanks{A. Lapidoth is with the
 Departement of Information Technology and Electrical Engineering, ETH Zurich,
 8092 Zurich, Switzerland.  Email:lapidoth@isi.ee.ethz.ch}}

\maketitle

\begin{abstract}
The two-user discrete memoryless state-dependent multiple-access 
channel (MAC) models a scenario in which two encoders transmit 
independent messages to a single receiver via a MAC whose 
channel law is governed by the pair of encoders' inputs and by an i.i.d. 
state random variable. 
In the cooperative state-dependent MAC model it is further assumed that 
Message~1 is shared by both encoders whereas Message~2
is known only to Encoder~2 -- the cognitive transmitter. 
The capacity of the cooperative state-dependent MAC where the realization
of the state sequence is known non-causally to the cognitive encoder
has been derived by Somekh-Baruch~\emph{et.~al.}.

In this work we dispense of the assumption that Message~1 is shared
a-priori by both encoders. Instead, we study the case in which
Encoder~2 cribs causally from Encoder~1. 
We determine the capacity region for both, the case
where Encoder~2 cribs strictly causal and the case where Encoder~2 
cribs causally from Encoder~1. 
\end{abstract}

\begin{keywords}
State-dependent MAC, Gel'fand-Pinsker channel, cribbing encoder.
\end{keywords}

\section{Introduction}

The two-user discrete memoryless state-dependent multiple-access 
channel (MAC) models a scenario in which two encoders transmit 
independent messages to a single receiver via a MAC whose 
channel law is governed by the pair of encoders' inputs and by an i.i.d. 
state random variable $S$. 
In the cooperative state-dependent MAC model it is further assumed that 
Message~1 is shared by both encoders whereas Message~2
is known only to Encoder~2 -- the cognitive transmitter. 
The capacity of the cooperative state-dependent MAC where the realization
of the state sequence is known non-causally to the cognitive encoder
has been derived by Somekh-Baruch~\emph{et.~al.} in \cite{anelia}.

In this work we dispense of the assumption that Message~1 is shared
a-priori by both encoders. Instead, we study a ``more realistic''
model in which Encoder~2 ``cribs''  and learns the sequence
of channel inputs emitted by Encoder~1 before 
generating its next channel input. Specifically, we study both, the case
where Encoder~2 cribs strictly causal -- i.e. its current channel input depends 
on its message as well as the past inputs of Encoder~1 (in
the sense of \cite[Situation~2]{frans}), and the case where Encoder~2 
cribs causally -- i.e. its current channel input depends 
on its message as well as all past including the current inputs of Encoder~1 
(in the sense of \cite[Situation~3]{frans}).
The model is depicted in Figure~1.

For both cases -- strictly causal cribbing as well as causal cribbing -- 
we provide a complete characterization of the capacity region.

The paper is organized as follows. In Section~II we provide a formal
definition for the state-dependent MAC with a cribbing encoder. 
In Section~III we present our main
results, while Section~IV is devoted for the proofs.

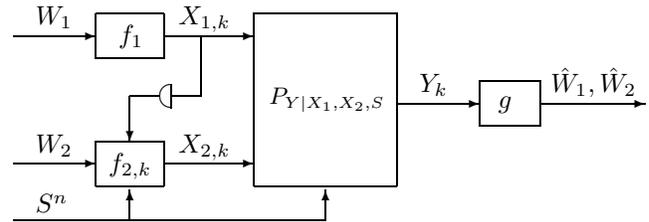
\begin{figure}
\centering
\setlength{\unitlength}{0.1mm}
\begin{picture}(850,320)
    \linethickness{0.1mm}
    \put(30,265){$W_1$}
    \put(0,250){\vector(1,0){110}}
    \put(110,220){\line(0,1){60}}
    \put(110,220){\line(1,0){90}}
    \put(110,280){\line(1,0){90}}
    \put(200,220){\line(0,1){60}}
    \put(140,240){$f_1$}
    \put(200,250){\vector(1,0){120}}
    \put(220,265){$X_{1,k}$}
    \put(250,250){\line(0,-1){80}}
    \put(210,170){\line(1,0){40}}
    \put(210,170){\oval(30,30)[l]}
    \put(210,155){\line(0,1){30}}
    \put(155,170){\line(1,0){40}}
    \put(155,170){\vector(0,-1){60}}
    \put(30,95){$W_2$}
    \put(0,80){\vector(1,0){110}}
    \put(110,50){\line(0,1){60}}
    \put(110,50){\line(1,0){90}}
    \put(110,110){\line(1,0){90}}
    \put(200,50){\line(0,1){60}}
    \put(127,70){$f_{2,k}$}
    \put(200,80){\vector(1,0){120}}
    \put(220,95){$X_{2,k}$}
    \put(340,160){\small $P_{Y|X_1,X_2,S}$}
    \put(320,50){\line(0,1){230}}
    \put(510,50){\line(0,1){230}}
    \put(320,280){\line(1,0){190}}
    \put(320,50){\line(1,0){190}}
    \put(510,160){\vector(1,0){110}}
    \put(540,175){$Y_k$}
    \put(0,5){\line(1,0){415}}
    \put(415,5){\vector(0,1){40}}
    \put(155,5){\vector(0,1){40}}
    \put(30,15){$S^n$}
    \put(620,130){\line(0,1){60}}
    \put(620,130){\line(1,0){80}}
    \put(620,190){\line(1,0){80}}
    \put(700,130){\line(0,1){60}}
    \put(645,155){$g$}
    \put(700,160){\vector(1,0){140}}
    \put(715,175){${\hat{W}}_1,\hat{W}_2$}
\end{picture}
\label{fig:setup} \caption{State-dependent MAC with a cribbing encoder.}
\end{figure}

\section{Channel model}

A discrete memoryless state-dependent multiple-access channel is a triple $\left( {\cal
X}_1\times{\cal X}_2\times{\cal S}, p(y|x_1,x_2,s),{\cal Y}\right)$
where ${\cal X}_1$ and ${\cal X}_2$ are finite sets corresponding to
the input alphabets of Encoder~1 and Encoder~2 respectively, ${\cal S}$ is a finite set
corresponding to the alphabet of the state governing the channel law, the
finite set ${\cal Y}$ is the output alphabet at the
receiver, and
$p(\cdot|x_1,x_2,s)$ is a collection of probability laws on
${\cal Y}$ indexed by the input symbols $x_1\in{\cal
X}_1$ and  $x_2\in{\cal X}_2$ and $s\in{\cal S}$. The channel's law extends to
$n$-tuples according to the memoryless law
\begin{IEEEeqnarray*}{l}
\Pr(y^n|x_{1}^{n},x_{2}^n,s^{n})=
 \prod_{k=1}^n p(y_k|x_{1,k},x_{2,k},s_k) \ ,
\end{IEEEeqnarray*}
where $x_{1,k},x_{2,k},s_{k}$ and $y_k$ denote the inputs, state and
output of the channel at time $k$, and $x_1^k\triangleq
(x_{1,1},\ldots,x_{1,k})$.

Encoder~1 sends a message $W_1$, which is drawn uniformly over the
set $\{1,\ldots,e^{nR_1}\}\triangleq {\cal W}_1$ to the receiver, while
Encoder~2 sends to the receiver a message $W_2$ which is independent of $W_1$ and 
is drawn uniformly over the set $\{1,\ldots,e^{nR_2}\}\triangleq {\cal W}_2$. 
The channel state sequence $S^n$, which is drawn i.i.d. according 
to the law $p_S$, is available non-causally to Encoder~2.
It is further assumed that Encoder~2 ``cribs'' causally and learns the sequence
of channel inputs emitted by Encoder~1 in all past transmissions (in
the sense of \cite[Situation~2]{frans}) before generating its next
channel input. The model is depicted in Figure~1.

An $(e^{nR_1},e^{nR_2},n)$ code for the state-dependent multiple-access channel with a
{\it strictly causal} cribbing encoder consists of:

1) \ Encoder~1 defined by a deterministic mapping
\begin{IEEEeqnarray}{l}
f_1 \colon  {\cal W}_1 \to {\cal X}_1^n
\label{eq:enc1}
\end{IEEEeqnarray}
which maps a message $w_1\in {\cal W}_1$ to a codeword $x_1^n\in
{\cal X}_1^n$.

2) \ Encoder~2 defined by a collection of encoding functions
\begin{IEEEeqnarray}{l}
f_{2,k}^{(sc)} \colon {\cal W}_2\times{\cal S}^n\times{\cal X}_1^{k-1}\to {\cal X}_2
 \ \ \ k=1,2,\ldots,n
\label{eq:enc2}
\end{IEEEeqnarray}
which, based on the message $w_2\in {\cal W}_2$, the state sequence $s^n\in {\cal S}^n$ and what was learned from the other encoder by cribbing $x_1^{k-1}\in {\cal X}_1^{k-1}$,
map into the next channel input $x_{2,k}\in {\cal X}_2$.

3) \  The receiver decoder defined by a mapping
\begin{IEEEeqnarray*}{l}
g \colon {\cal Y}^n \to {\cal W}_1\times{\cal W}_2
\end{IEEEeqnarray*}
which maps a received sequence $y^n$ to a message pair
$({\hat{w}}_1,\hat{w}_2)\in {\cal W}_1\times{\cal W}_2$.

\medskip

An $(e^{nR_1},e^{nR_2},n)$ code for the state-dependent multiple-access channel with a
{\it causal} cribbing encoder differs from that for a strictly causal encoder just in the encoding rule at Encoder~2 which is defined by a collection of encoding functions
\begin{IEEEeqnarray}{l}
f_{2,k}^{(c)} \colon {\cal W}_2\times{\cal S}^n\times{\cal X}_1^{k}\to {\cal X}_2
 \ \ \ k=1,2,\ldots,n
\label{eq:enc2c}
\end{IEEEeqnarray}
which, based on the message $w_2\in {\cal W}_2$, the state sequence $s^n\in {\cal S}^n$ and what was learned from the other encoder by cribbing $x_1^{k}\in {\cal X}_1^{k}$,
map into the current channel input $x_{2,k}\in {\cal X}_2$.

\medskip

For a given code, the block average probability of error is
\begin{IEEEeqnarray*}{l}
P_e^{(n)}= \frac{1}{e^{n(R_1+R_2)}}
  \sum_{w_1=1}^{e^{nR_1}}\sum_{w_2=1}^{e^{nR_2}}
 P_e^{(n)}(w_1,w_2)
\end{IEEEeqnarray*}
where
\begin{IEEEeqnarray*}{l}
P_e^{(n)}(w_1,w_2)= \nonumber \\
 \ \ \ \ \ \ \
 \Pr\left\{(\hat{w}_1,\hat{w}_2)\neq (w_1,w_2) | (w_1,w_2) \ \mbox{sent}
 \right\}.
\end{IEEEeqnarray*}
A rate-pair $(R_1,R_2)$ is said to be achievable if
there exists a sequence of $(e^{nR_1},e^{nR_2},n)$ codes with
$\lim_{n\to\infty} P_e^{(n)}=0$. The capacity region of the state-dependent MAC with a
cribbing encoder is the closure of the set of achievable rate-pairs.

\section{Main results}

Our first result is a characterization of the capacity region for the two-user discrete memoryless 
state-dependent MAC with state-sequence available non-causally at a strictly causal cribbing encoder. By combining the coding strategies from \cite{anelia} and \cite{frans} we prove the
following.

\vskip.1truein

\begin{theorem} \label{th:thm1}
Consider the discrete memoryless state-dependent MAC 
$\left( {\cal X}_1\times{\cal X}_2\times{\cal S}, p(y|x_1,x_2,s),{\cal Y}\right)$
with state-sequence available non-causally at a strictly causal cribbing encoder and finite
alphabets ${\cal S},{\cal X}_1,{\cal X}_2$. 
The capacity region of this channel is 
\begin{IEEEeqnarray}{l}
{\cal C}=\bigcup_{p_{VSUX_{1}X_{2}Y}} 
\biggl\{ (R_1,R_2): \nonumber \\
0\leq R_1\leq  H(X_{1}|V) \nonumber \\
0\leq R_2\leq I(U;Y|VX_1)-I(U;S|V) \nonumber \\
0\leq R_1+R_2\leq  I(VUX_1;Y)-I(U;S|V)\biggr\} ,
\label{eq:r11}
\end{IEEEeqnarray}
where the union in (\ref{eq:r11}) is over all laws on
$V\in\set{V},S\in \set{S},U\in\set{U}, X_1\in \set{X}_1, X_2\in\set{X}_2,
Y\in\set{Y}$ of the form 
\begin{IEEEeqnarray}{l}
p_{VSUX_{1}X_{2}Y}(v,s,u,x_{1},x_{2},y) \nonumber \\
 =  p_{V}(v)p_S(s)p_{X_{1}|V}(x_{1}|v)
  p_{UX_{2}|SV}(u,x_{2}|s,v)p(y|x_{1},x_{2},s) . \nonumber \\
  \label{eq:jointlawp}
\end{IEEEeqnarray}
The cardinalities of the auxiliary 
random variables $V$ and $U$ are bounded by 
\begin{IEEEeqnarray*}{l}
|{\cal V}| \leq |{\cal X}_1| |{\cal X}_2| |{\cal S}|  + 5  \nonumber \\
|{\cal U}\|\leq |{\cal X}_1| |{\cal X}_2| |{\cal S}| |{\cal V}| + 2.
\end{IEEEeqnarray*}

\end{theorem}

\vskip.1truein

Our second result is a characterization of the capacity region for the two-user discrete memoryless 
state-dependent MAC with state-sequence available non-causally at a causal cribbing encoder.

\vskip.1truein

\begin{theorem} \label{th:thm2}
Consider the discrete memoryless state-dependent MAC 
$\left( {\cal X}_1\times{\cal X}_2\times{\cal S}, p(y|x_1,x_2,s),{\cal Y}\right)$
with state-sequence available non-causally at a causal cribbing encoder and finite
alphabets ${\cal S},{\cal X}_1,{\cal X}_2$. 
The capacity region of this channel is the set of rate pairs satisfying (\ref{eq:r11}) 
except that the union is taken over all laws on
$V\in\set{V},S\in \set{S},U\in\set{U}, X_1\in \set{X}_1, X_2\in\set{X}_2,
Y\in\set{Y}$ of the form 
\begin{IEEEeqnarray}{l}
p_{VSUX_{1}X_{2}Y}(v,s,u,x_{1},x_{2},y) \nonumber \\
 =  p_{V}(v)p_S(s)p_{X_{1}|V}(x_{1}|v)p_{U|SV}(u|s,v) \nonumber \\
 \ \ \ \quad p_{X_{2}|VUSX_1}(x_{2}|v,u,s,x_1)p(y|x_{1},x_{2},s) . 
  \label{eq:jointlawpc}
\end{IEEEeqnarray}
\end{theorem}

\section{Proofs}

\subsection{Proof of the achievability part in Theorem~\ref{th:thm1}}

We propose a coding scheme that is based on Block-Markov
superposition encoding and which combines the coding technique of
\cite{anelia} with that of \cite{frans}, while the decoder uses backward 
decoding.

\subsubsection{Coding Scheme}\label{sec:coding-scheme}
We consider $B$ blocks, each of $n$ symbols. A sequence of
$B-1$ message pairs $(W_1^{(b)},W_2^{(b)})$, for $b=1,\ldots, B-1$, 
will be transmitted during $B$ transmission blocks. Here the
sequence $\{W_1^{(b)}\}$ is an i.i.d. sequence of uniform random
variables over $\left\{1,\ldots,e^{nR_1}\right\}$ and independent
thereof $\{W_2^{(b)}\}$ is an i.i.d. sequence of uniform random
variables over $\left\{1,\ldots,e^{nR_{2}}\right\}$. As $B\to\infty$, for fixed
$n$, the rate pair of the message $(W_1,W_2)$,
$(\tilde{R}_1,\tilde{R}_2)=(R_1(B-1)/B,R_2(B-1)/B)$, is
arbitrarily close to $(R_1,R_2)$.

We assume a tuple of random variables $V\in\set{V},S\in \set{S},U\in\set{U},
X_1\in \set{X}_1, X_2\in\set{X}_2, Y\in\set{Y},$ of
joint law (\ref{eq:jointlawp}).

\vskip.1truein

{\it Random coding and partitioning:} In each block
$b,b=1,2,\ldots,B$, we shall use the following code.
\begin{itemize}
\item Generate $e^{nR_1}$ sequences
  $\makebox{{\boldmath $v$}}=({v}_{1},\ldots,{v}_{n})$,
  each with probability
  $\Pr\left({\mbox{\boldmath $v$}}\right)=\prod_{k=1}^{n}p_{V}({v}_{k})$.
   Label them
  ${\mbox{\boldmath $v$}}\left(\omega_0\right)$ where
  $\omega_{0}\in\left\{1,\ldots,e^{nR_{1}}\right\}$.

\item For each ${\mbox{\boldmath $v$}}\left(\omega_0\right)$
  generate $e^{nR_1}$ sequences
  ${\mbox{\boldmath $x$}_1}=({x}_{1,1},{x}_{1,2},\ldots,{x}_{1,n})$,
  each with probability
  $\Pr\left({\mbox{\boldmath $x$}_1}|{\mbox{\boldmath
  $v$}}\left(\omega_0\right)\right)=
  \prod_{k=1}^{n}p_{X_1|V}({x}_{1,k}|v_k(\omega_0))$. Label them
  ${\mbox{\boldmath $x$}_1}\left(i,\omega_0\right), i\in
\left\{1,\ldots,e^{nR_{1}}\right\}$.

\item For each ${\mbox{\boldmath $v$}}\left(\omega_0\right)$
  generate $e^{n(R_{2}+R')}$ sequences
  ${\mbox{\boldmath $u$}}=({u}_{1},{u}_{2},\ldots,{u}_{n})$,
  each with probability
  $\Pr\left({\mbox{\boldmath $u$}}|{\mbox{\boldmath
  $v$}}\left(\omega_0\right)\right)=
  \prod_{k=1}^{n}p_{U|V}({u}_{k}|v_k(\omega_0))$. 
   Randomly partition the set
  $\left\{{\mbox{\boldmath $u$}}\right\}$ into $e^{nR_2}$ bins, each 
  consisting of $e^{nR'}$ codewords. Now 
  label the codewords by 
  ${\mbox{\boldmath $u$}}\left(j,\jmath,\omega_0\right), j\in
  \{1,\ldots,e^{nR_{2}}\},  \jmath \in
  \{1,\ldots,e^{nR'}\}$ where $j$ identifies the bin and $\jmath$ 
  the index within the bin.
\end{itemize}

\vskip.1truein

{\it Encoding :} We denote the realizations of the sequences
$\{W_1^{(b)}\}$ and $\{W_{2}^{(b)}\}$ by
$\{w_1^{(b)}\}$ and $\{w_{2}^{(b)}\}$, and the realization of
the state sequence $(S_1^{(b)},S_2^{(b)},\ldots,S_n^{(b)})$ by
${\mbox{\boldmath $s$}}^{(b)}$.  The
code builds upon a Block-Markov structure in which the message
$(w_1^{(b)},w_{2}^{(b)})$ is encoded over the
successive blocks $b$ and $(b+1)$ such that,
$\omega_0^{(b+1)}=w_1^{(b)}$, for $b=1,\ldots, B-1$.

\medskip

The messages $\{w_1^{(b)}\}$ and
$\{w_{2}^{(b)}\}$, $b=1,2,\ldots,B-1$ are encoded as follows:

In block $1$ the encoders send
\begin{IEEEeqnarray*}{l}
\mbox{\boldmath $x$}_1^{(1)} = \mbox{\boldmath $x$}_1(w_1^{(1)},1)
 \nonumber \\
\mbox{\boldmath $x$}_2^{(1)} = \mbox{\boldmath
$x$}_2({\mbox{\boldmath $s$}}^{(1)},w_{2}^{(1)},1).
\end{IEEEeqnarray*}
Here, the encoding 
$\mbox{\boldmath $x$}_2({\mbox{\boldmath $s$}}^{(b)},w_{2}^{(b)},\omega_0^{(b)})$ 
is defined as follows:
\begin{enumerate}
\item Find the typical ${\mbox{\boldmath $u$}}(w_{2}^{(b)},\jmath_0,\omega_0^{(b)})$: 
Search within the bin 
${\mbox{\boldmath $u$}}(w_{2}^{(b)},\cdot,\omega_0^{(b)})$ 
 for the lowest $\jmath_0\in\{1,\ldots,e^{nR'}\}$ such that 
 ${\mbox{\boldmath $u$}}(w_{2}^{(b)},\jmath_0,\omega_0^{(b)})$ is jointly typical with the pair
 $({\mbox{\boldmath $v$}}(\omega_0^{(b)}),{\mbox{\boldmath $s$}}^{(b)})$; denote this 
 $\jmath_0$ as $\jmath_0({\mbox{\boldmath $s$}}^{(b)},w_{2}^{(b)},\omega_0^{(b)})$. If
 such $\jmath_0$ is not found or if the state sequence ${\mbox{\boldmath $s$}}^{(b)}$ is
 non-typical an error is declared and
 $\jmath_0({\mbox{\boldmath $s$}}^{(b)},w_{2}^{(b)},\omega_0^{(b)})=1$.   

\item Generate the codeword 
$\mbox{\boldmath $x$}_2({\mbox{\boldmath $s$}}^{(b)},w_{2}^{(b)},\omega_0^{(b)})$ by
drawing its components i.i.d. conditionally on the triple $({\mbox{\boldmath $s$}}^{(b)},{\mbox{\boldmath $u$}}(w_{2}^{(b)},\jmath_0,\omega_0^{(b)}),{\mbox{\boldmath $v$}}(\omega_0^{(b)}))$,
where the conditional law is induced by (\ref{eq:jointlawp}).  

\end{enumerate}

Suppose that, as a result of cribbing from Encoder~1, before the
beginning of block $b=2,3,\ldots,B$, Encoder~2 has an estimate
$\hat{\hat{w}}_1^{(b-1)}$ for $w_1^{(b-1)}$. Then, in block
$b=2,3,\ldots,B-1$, the encoders send
\begin{IEEEeqnarray*}{l}
\mbox{\boldmath $x$}_1^{(b)} = \mbox{\boldmath
$x$}_1(w_1^{(b)},w_1^{(b-1)})
 \nonumber \\
\mbox{\boldmath $x$}_2^{(b)} =
 \mbox{\boldmath $x$}_2({\mbox{\boldmath $s$}}^{(b)},w_{2}^{(b)},\hat{\hat{w}}_1^{(b-1)}),
\end{IEEEeqnarray*}
and in block $B$
\begin{IEEEeqnarray*}{l}
\mbox{\boldmath $x$}_1^{(B)} = \mbox{\boldmath $x$}_1(1,w_1^{(B-1)})
 \nonumber \\
\mbox{\boldmath $x$}_2^{(B)} = \mbox{\boldmath
$x$}_2({\mbox{\boldmath $s$}}^{(B)},1,\hat{\hat{w}}_1^{(B-1)}).
\end{IEEEeqnarray*}

\vskip.1truein

{\it Decoding at the receiver:} After the reception of block-$B$
the receiver uses backward decoding starting from block $B$
downward to block $1$ and decodes the messages as follows.

In block $B$ the receiver looks for $\hat{w}_1^{(B-1)}$ such that
\begin{IEEEeqnarray*}{l}
\left( \mbox{\boldmath $v$}({\hat{w}}_{1}^{(B-1)}), \mbox{\boldmath
$x$}_1(1,{\hat{w}}_{1}^{(B-1)}),
{\mbox{\boldmath $u$}}(1,\jmath_0,{\hat{w}}_{1}^{(B-1)}), \right.
 \nonumber \\
\hspace{1.3cm} \left. \mbox{\boldmath
$x$}_2({\mbox{\boldmath $s$}}^{(B)},1,{\hat{w}}_{1}^{(B-1)}),
 \mbox{\boldmath $y$}^{(B)} \right) \in{\cal A}_{\epsilon}(V,X_1,U,X_2,Y),
\end{IEEEeqnarray*}
where $\jmath_0= \jmath_0({\mbox{\boldmath $s$}}^{(B)},1,{\hat{w}}_{1}^{(B-1)})$.

Next, assume that, decoding backwards up to (and including) block
$b+1$, the receiver decoded $\hat{w}_1^{(B-1)},(\hat{w}_{2}^{(B-1)},\hat{w}_1^{(B-2)}),
\ldots,(\hat{w}_2^{(b+1)},\hat{w}_1^{(b)})$. 
To decode block $b$, the receiver looks for $(\hat{w}_2^{(b)},\hat{w}_1^{(b-1)})$ such that
\begin{IEEEeqnarray*}{l}
\left( \mbox{\boldmath $v$}({\hat{w}}_{1}^{(b-1)}), \mbox{\boldmath
$x$}_1(\hat{w}_1^{(b)},{\hat{w}}_{1}^{(b-1)}),
{\mbox{\boldmath $u$}}(\hat{w}_2^{(b)},\jmath_0,{\hat{w}}_{1}^{(b-1)}), \right.
 \nonumber \\
\hspace{1.3cm} \left. \mbox{\boldmath
$x$}_2({\mbox{\boldmath $s$}}^{(b)},\hat{w}_2^{(b)},{\hat{w}}_{1}^{(b-1)}),
 \mbox{\boldmath $y$}^{(b)} \right) \in{\cal A}_{\epsilon}(V,X_1,U,X_2,Y),
\end{IEEEeqnarray*}
where $\jmath_0= \jmath_0({\mbox{\boldmath $s$}}^{(b)},\hat{w}_{2}^{(b)},{\hat{w}}_{1}^{(b-1)})$.

\medskip

{\it Decoding at Encoder~2:} To obtain cooperation, after block
$b=1,2,\ldots,B-1$, Encoder~2 chooses $\tilde{w}_1^{(b)}$ such that
\begin{IEEEeqnarray*}{l}
\left( \mbox{\boldmath $v$}(\tilde{\omega}_{0}^{(b)}),
\mbox{\boldmath
$x$}_1(\tilde{w}_{1}^{(b)},\tilde{\omega}_{0}^{(b)}),
\mbox{\boldmath $x$}_1^{(b)} \right)
 \in{\cal A}_{\epsilon}(V,X_1,X_1),
\end{IEEEeqnarray*}
where $\tilde{\omega}_{0}^{(b)}=\tilde{w}_{1}^{(b-1)}$ was
determined at the end of block $b-1$ and
$\tilde{\omega}_{0}^{(1)}=1$.

When a decoding step either fails to recover a unique index (or
index pair) which satisfies the decoding rule, or there is more
than one index (or index pair), then an index (or an index pair)
is chosen at random.

\vskip.1truein

\subsubsection{Bounding the Probability of Error}

Genie-aided arguments as in \cite{rimoldiurbanke96} and
\cite{wozencraftjacobs65} can be used to show that the probability
that either Endoder~2 makes an encoding error or the receiver 
makes a decoding error after block $b$ in the
above scheme is upper bounded by the probability that at least one
of the following events $E_0^{(b)}-E_5^{(b)}$ happens.

{\it Error events:}

\begin{itemize}
\item $E_0^{(b)}:$
\begin{IEEEeqnarray*}{l}
\left(\mbox{\boldmath $v$}(\omega_{0}^{(b)}),\mbox{\boldmath
$u$}(w_{2}^{(b)},\jmath_0,\omega_{0}^{(b)}), \mbox{\boldmath
$x$}_{1}(w_1^{(b)},\omega_{0}^{(b)}) \right) \\
\hspace{2.5cm}  \not\in {\cal A}_{\epsilon}(V,U,X_{1}).
\IEEEeqnarraynumspace
\end{IEEEeqnarray*}

\item $E_1^{(b)}$: There exists $\tilde{w}_1\neq
w_{1}^{(b)}$ such that
\begin{IEEEeqnarray*}{l}
\left( \mbox{\boldmath $v$}({\omega}_{0}^{(b)}), \mbox{\boldmath
$x$}_1(\tilde{w}_{1},{\omega}_{0}^{(b)}), \mbox{\boldmath
$x$}_1^{(b)} \right)
 \in{\cal A}_{\epsilon}(V,X_1,X_1).
\end{IEEEeqnarray*}

\item $E_2^{(b)}:$ There doesn't exist $\jmath_0\in\{1,\ldots,e^{nR'}\}$ such
that
\begin{IEEEeqnarray*}{l}
\left(\mbox{\boldmath $v$}(\omega_{0}^{(b)}),\mbox{\boldmath
$u$}(w_{2}^{(b)},\jmath_0,\omega_{0}^{(b)}), {\mbox{\boldmath $s$}}^{(b)} \right) 
\in {\cal A}_{\epsilon}(V,U,S).
\IEEEeqnarraynumspace
\end{IEEEeqnarray*}

\item $E_3^{(b)}:$
\begin{IEEEeqnarray*}{l}
\left(\mbox{\boldmath $v$}(\omega_{0}^{(b)}),\mbox{\boldmath
$u$}(w_{2}^{(b)},\jmath_0,\omega_{0}^{(b)}), \mbox{\boldmath
$x$}_{1}(w_1^{(b)},\omega_{0}^{(b)}),\right. \\
 \left. \hspace{1.0cm}  \mbox{\boldmath
$x$}_2({\mbox{\boldmath $s$}}^{(b)},{w}_2^{(b)},{{w}}_{1}^{(b-1)}),
\mbox{\boldmath $y$}^{(b)} \right) \\
\hspace{2.5cm}  \not\in {\cal A}_{\epsilon}(V,U,X_{1},X_{2},Y).
\IEEEeqnarraynumspace
\end{IEEEeqnarray*}

\item $E_4^{(b)}$: There exists $\tilde{\omega}_0\neq \omega_0^{(b)}$ such that
\begin{IEEEeqnarray*}{l}
\left( \mbox{\boldmath $v$}(\tilde{\omega}_{0}^{(b)}),
\mbox{\boldmath $x$}_1(w_1^{(b)},\tilde{\omega}_{0}^{(b)}),
\mbox{\boldmath $u$}(j,\jmath_0,\tilde{\omega}_{0}^{(b)}), \right.
\nonumber \\
\left. \hspace{1.0cm} \mbox{\boldmath
$x$}_2({\mbox{\boldmath $s$}}^{(b)},j,\tilde{\omega}_0^{(b)}),
 \mbox{\boldmath $y$}^{(b)} \right) \hspace{1cm}\nonumber \\
\hspace{2.5cm}\in{\cal A}_{\epsilon}(V,U,X_1,X_2,Y),
\end{IEEEeqnarray*}
for some pair $(j,\jmath_0) \ , \ j\in{\cal W}_{2} \ , \ \jmath_0\in\{1,\ldots,e^{nR'}\}$.

\item $E_5^{(b)}$: There exists $\tilde{w}_{2}\neq w_{2}^{(b)}$ such that
\begin{IEEEeqnarray*}{l}
\left( \mbox{\boldmath $v$}({\omega}_{0}^{(b)}),
\mbox{\boldmath $x$}_1(w_1^{(b)},{\omega}_{0}^{(b)}),
\mbox{\boldmath $u$}(\tilde{w}_2,\jmath_0,{\omega}_{0}^{(b)}), \right.
\nonumber \\
\left. \hspace{1.0cm} \mbox{\boldmath
$x$}_2({\mbox{\boldmath $s$}}^{(b)},\tilde{w}_2,{\omega}_0^{(b)}),
 \mbox{\boldmath $y$}^{(b)} \right) \hspace{1cm}\nonumber \\
\hspace{2.5cm}\in{\cal A}_{\epsilon}(V,U,X_1,X_2,Y),
\end{IEEEeqnarray*}
for some index $\jmath_0\in\{1,\ldots,e^{nR'}\}$.

\end{itemize}

We define the event
\begin{IEEEeqnarray*}{rCl}
F_1^{(b)} \triangleq \bigcup_{j=4}^{5}E_{j}^{(b)}, \qquad
b=1,\ldots, B,
\end{IEEEeqnarray*}
the event
\begin{IEEEeqnarray*}{rCl}
F_2 \triangleq \bigcup_{j=1}^{B}E_{0}^{(b)},
\end{IEEEeqnarray*}
the event
\begin{IEEEeqnarray*}{rCl}
F_3 \triangleq \bigcup_{j=1}^{B} \left(E_{0}^{(b)}\cup
E_{1}^{(b)}\right),
\end{IEEEeqnarray*}
the event
\begin{IEEEeqnarray*}{rCl}
F_4 \triangleq \bigcup_{j=1}^{B} \left(E_{0}^{(b)}\cup
E_{1}^{(b)}\cup E_2^{(b)}\right),
\end{IEEEeqnarray*}
and the event
\begin{IEEEeqnarray*}{rCl}
F_4 \triangleq \bigcup_{j=1}^{B} \left(E_{0}^{(b)}\cup
E_{1}^{(b)}\cup E_2^{(b)}\cup E_3^{(b)}\right).
\end{IEEEeqnarray*}

We can upper bound the average probability of error
$\bar{P}_e$ averaged over all codebooks and all random partitions by
\begin{IEEEeqnarray*}{rCl}\label{eq:Petot}
\bar{P}_{e} & \leq &  \sum_{b=1}^{B}\left\{ \Pr\left[E_{0}^{(b)}\right] +
\Pr\left[E_{1}^{(b)}|{F_2}^{c},{E_{1}^{(1\ldots
 b-1)}}^c\right] \right\} \nonumber \\
& & +  \sum_{b=1}^{B}\left\{ \Pr\left[E_{2}^{(b)}|F_3^{c}\right] +
\Pr\left[E_{3}^{(b)}|{F_4}^{c},{E_{3}^{(1\ldots
 b-1)}}^c\right] \right\} \nonumber \\
& & +\sum_{b=1}^{B}\Pr\left[F_1^{(b)}|{F_4}^{c},{F_1^{(b+1\ldots
B)}}^{c}\right],
\end{IEEEeqnarray*}
where $F{^{(1\ldots b-1)}}^{c}$ denotes the complement of the event
$F^{(1)} \cup \ldots \cup F^{(b-1)}$.

Furthermore, we can upper bound each of the summands in the last
component as
\begin{IEEEeqnarray*}{rCl}\lefteqn{
\Pr\left(F_1^{(b)}|{F_4}^{c},{F_1^{(b+1\ldots
B)}}^{c}\right)}\\
& =& \Pr\left(\bigcup^{5}_{j=4}E_{j}^{(b)}|
 {F_4}^{c},{F_1^{(b+1\ldots B)}}^{c}  \right)\\
 &\leq & \Pr\left(E_{4}^{(b)} \big|
   {F_4}^{c},{F_1^{(b+1\ldots B)}}^{c}\right) \\
 && + \Pr\left(E_{5}^{(b)} \big|
   {F_4}^{c},{F_1^{(b+1\ldots B)}}^{c}\right) .
\end{IEEEeqnarray*}

In the following we separately examine each of the above summands.

By Lemma~\ref{th:lm1} (Appendix A)  
$\Pr\left[E_{3}^{(b)}|{F_4}^{c},{E_{3}^{(1\ldots b-1)}}^c\right] $ and
$\Pr\left[E_{0}^{(b)}\right]$ can be
made arbitrarily small for sufficiently large $n$.

Also, by Lemma~\ref{th:lm2}:
\begin{itemize}
\item  If
\begin{eqnarray}
R_1 < H(X_1|V), \label{eq:decenc2}
\end{eqnarray}
then $\Pr\left[E_{1}^{(b)}|{F_2}^{c},{E_{1}^{(1\ldots
 b-1)}}^c\right]$ can be made arbitrarily small, provided that $n$ is
sufficiently large;
\item  If
\begin{eqnarray}
R_1+R_{2}+R' < I(VUX_1;Y), \label{eq:decrec1}
\end{eqnarray}
then $\Pr\left(E_{4}^{(b)} \big|
   {F_4}^{c},{F_1^{(b+1\ldots B)}}^{c}\right)$
can be made arbitrarily small, provided that $n$ is sufficiently
large;
\item If
\begin{eqnarray} 
R_2+R'<I(U;Y|VX_1) \label{eq:decrec21}
\end{eqnarray}
then $\Pr\left(E_{5}^{(b)} \big|
   {F_4}^{c},{F_1^{(b+1\ldots B)}}^{c}\right)$
can be made arbitrarily small, provided that $n$ is sufficiently
large;
\end{itemize}
 Finally, by the covering lemma (See \cite{ahlkor,wy1,ber} or
\cite[Chapter 13]{cov}), if
\begin{eqnarray}
R'>I(U;S|V) \label{eq:decrec22}
\end{eqnarray}
then $\Pr\left[E_{2}^{(b)} \big|{F_3}^{c}\right]$
can be made arbitrarily small, provided that $n$ is sufficiently
large.

The combination of (\ref{eq:decenc2}), (\ref{eq:decrec1}),
(\ref{eq:decrec21}), and (\ref{eq:decrec22}) establishes the achievability of the
rate region (\ref{eq:r11})  for a law of the form (\ref{eq:jointlawp}).

\vskip.1truein

\subsection{Proof of the converse in Theorem~\ref{th:thm1}}

Consider an $(e^{nR_1},e^{nR_2},n)$ code with average block error
probability $P_e^{(n)}$, and a law on ${\cal W}_1\times{\cal
W}_2\times{\cal X}_1^n\times{\cal X}_2^n\times{\cal Y}^n\times{\cal
S}^n$ given by
\begin{IEEEeqnarray}{l}
p_{W_1W_2X_1^nX_2^nS^nY^n} \nonumber \\
=p_{W_1}p_{W_2}I_{\{X_1^n=f_1(W_1)\}}p_{X_2^n|W_1W_2S^n}
\prod_{k=1}^np_{Y_k|X_{1,k}X_{2,k}S_k}. \nonumber \\
\label{eq:convlaw}
\end{IEEEeqnarray}

\medskip

Let $V_k$ be the random variable defined by
\begin{IEEEeqnarray}{rCl}
 V_k & \triangleq & X_1^{k-1},
\label{eq:defVk}
\end{IEEEeqnarray}
and let $U_k$ be the random variable defined by
\begin{IEEEeqnarray}{rCl}
 U_k & \triangleq & W_2Y^{k-1}S_{k+1}^{n}.
\label{eq:defUk}
\end{IEEEeqnarray}

We start with an upper bound on $R_1$ by following similar steps as in
\cite[Section~V---Converse for situation 2]{frans}.
\begin{IEEEeqnarray}{rCl}
nR_1 & = & H(W_1|W_2) \nonumber\\
 & = & I(W_1;Y^n|W_2)+H(W_1|W_2Y^n) \nonumber \\
 & \leq & I(W_1;Y^n|W_2)+n\delta(P_e) \nonumber \\
 & \stackrel{(a)}{=} & I(X_1^n;Y^n|W_2)+n\delta(P_e) \nonumber \\
 & = & \sum_{k=1}^nI(X_{1,k};Y^n|W_2X_1^{k-1}) + n\delta(P_e) \nonumber \\
 & \leq & \sum_{k=1}^n H(X_{1,k}|X_1^{k-1}) + n\delta(P_e) \nonumber \\
 & = & \sum_{k=1}^n H(X_{1,k}|V_{k}) + n\delta(P_e).
\label{eq:R1upr1}
\end{IEEEeqnarray}
where $(a)$ follows from the encoding relation (\ref{eq:enc1}).

Next, consider $R_2$
\begin{IEEEeqnarray}{rCl}
nR_2 & = & H(W_2|W_1) \nonumber\\
 & \leq & I(W_2;Y^n|W_1)+n\delta(P_e) \nonumber \\
 & = & \sum_{k=1}^n I(W_2;Y_k|W_1Y^{k-1})+n\delta(P_e) \nonumber \\
 & \leq & \sum_{k=1}^n I(W_2Y^{k-1};Y_k|W_1)+n\delta(P_e) \nonumber \\
 & = & \sum_{k=1}^n \left[I(W_2Y^{k-1}S_{k+1}^n;Y_k|W_1) \right. \nonumber \\
  & & \quad \left. -I(S_{k+1}^n;Y_k|W_1W_2Y^{k-1})\right]+n\delta(P_e) \nonumber \\
 & \stackrel{(b)}{=} & \sum_{k=1}^n\left[ I(W_2Y^{k-1}S_{k+1}^n;Y_k|W_1) \right. \nonumber \\
  & & \quad \left. -I(Y^{k-1};S_k|W_1W_2S_{k+1}^n)\right]+n\delta(P_e) \nonumber \\
 & \stackrel{(c)}{=} & \sum_{k=1}^n \left[ I(W_2Y^{k-1}S_{k+1}^n;Y_k|W_1) \right. \nonumber \\
  & & \quad \left. -I(W_2Y^{k-1}S_{k+1}^n;S_k|W_1)\right]+n\delta(P_e) \nonumber \\
 & \stackrel{(d)}{=}  & \sum_{k=1}^n \left[ I(W_2Y^{k-1}S_{k+1}^n;Y_k|W_1X_1^{k-1}X_{1,k})
   \right. \nonumber \\
  & & \quad \left. -I(W_2Y^{k-1}S_{k+1}^n;S_k|W_1X_1^{k-1})\right]+n\delta(P_e) \nonumber \\
 & \stackrel{(e)}{=}  & \sum_{k=1}^n \left[ I(W_2Y^{k-1}S_{k+1}^n;Y_k|X_1^{k-1}X_{1,k})
   \right. \nonumber \\
  & & \quad \left. -I(W_2Y^{k-1}S_{k+1}^n;S_k|W_1X_1^{k-1})\right] +n\delta(P_e) \nonumber \\
 & \stackrel{(f)}{=}  & \sum_{k=1}^n \left[ I(W_2Y^{k-1}S_{k+1}^n;Y_k|X_1^{k-1}X_{1,k})
   \right. \nonumber \\
  & & \quad \left. -I(W_2Y^{k-1}S_{k+1}^n;S_k|X_1^{k-1})\right]+n\delta(P_e) \nonumber \\
 & = & \sum_{k=1}^n \left[ I(U_k;Y_k|V_{k}X_{1,k})-I(U_k;S_k|V_k)\right] .
\label{eq:R2upr1}
\end{IEEEeqnarray}
Here,
\begin{itemize}
\item[$(b)$] follows by the Csisz\'{a}r-K\"{o}rner's identity \cite[Lemma 7]{csi};
\item[$(c)$] follows since $(W_2S_{k+1}^n)$ is independent of $S_k$;
\item[$(d)$] follows by the encoding relation (\ref{eq:enc1});
\item[$(e)$] follows since  
  $W_1\Markov X_{1,k}X_1^{k-1}\Markov W_2Y_kY^{k-1}S_{k+1}^n$ and
  $W_1\Markov X_{1,k}X_1^{k-1}\Markov Y_k$ are Markov strings; and 
\item[$(f)$] follows since  
  $W_1\Markov X_1^{k-1}\Markov W_2S_kY^{k-1}S_{k+1}^n$ is a
  Markov string.
\end{itemize}

Finally, we consider the sum-rate $R_1+R_2$
\begin{IEEEeqnarray}{rCl}
n(R_1+R_2) & = & H(W_1W_2) \nonumber\\
 & \leq & I(W_1W_2;Y^n)+n\delta(P_e) \nonumber \\
 & = & \sum_{k=1}^n I(W_1W_2;Y_k|Y^{k-1})+n\delta(P_e) \nonumber \\
 & \stackrel{(g)}{\leq} & \sum_{k=1}^n \left[I(W_1W_2Y^{k-1}S_{k+1}^n;Y_k) \right.  \nonumber \\
  & & \quad \left. -I(W_1W_2Y^{k-1}S_{k+1}^n;S_k)\right]+n\delta(P_e) \nonumber \\
 & = & \sum_{k=1}^n \left[ I(W_1X_1^{k-1}X_{1,k}W_2Y^{k-1}S_{k+1}^n;Y_k)
   \right. \nonumber \\
  & & \quad \left. -I(W_1W_2Y^{k-1}S_{k+1}^n;S_k)\right]+n\delta(P_e) \nonumber \\
 & \stackrel{(h)}{=}  & \sum_{k=1}^n\left[ I(X_1^{k-1}X_{1,k}W_2Y^{k-1}S_{k+1}^n;Y_k)
   \right. \nonumber \\
  & & \quad \left. -I(W_1W_2Y^{k-1}S_{k+1}^n;S_k)\right]+n\delta(P_e) \nonumber \\
 & =  & \sum_{k=1}^n \left[ I(V_kU_kX_{1,k};Y_k)-I(W_1;S_k) \right.
   \nonumber \\
  & & \quad \left. -I(W_2Y^{k-1}S_{k+1}^n;S_k|W_1)\right]+n\delta(P_e) \nonumber \\
 & =  & \sum_{k=1}^n \left[ I(V_kU_kX_{1,k};Y_k)-I(W_1;S_k) \right.
   \nonumber \\
  & & \quad \left. -I(W_2Y^{k-1}S_{k+1}^n;S_k|W_1X_1^{k-1})\right]+n\delta(P_e) \nonumber \\
 & \stackrel{(i)}{=} & \sum_{k=1}^n \left[I(V_kU_kX_{1,k};Y_k)-I(U_k;S_k|V_k)\right] .
\label{eq:sumRupr1}
\end{IEEEeqnarray}
Here,
\begin{itemize}
\item[$(g)$] follows by the same procedure as $(b)$ and $(c)$;
\item[$(h)$] follows by the encoding relation (\ref{eq:enc1}) and since  
    $W_1\Markov X_{1,k}X_1^{k-1}W_2Y^{k-1}S_{k+1}^n\Markov Y_k$ is a Markov 
    string; and 
\item[$(i)$] follows since $W_1$ is independent of $S_k$ and since
  $W_1\Markov X_1^{k-1}W_2Y^{k-1}S_{k+1}^n \Markov S_k$ and
  $W_1\Markov X_1^{k-1}\Markov S_k$ are Markov strings. 
  
\end{itemize}

\medskip

Next we verify the joint law of the auxiliary random variables.

By \eqref{eq:convlaw} and the encoding rule (\ref{eq:enc2}) we may write
\begin{IEEEeqnarray*}{l}
p_{W_1W_2X_1^{k-1}X_{1,k}S^{k-1}S_kS_{k+1}^nX_2^kY^{k-1}}= \nonumber \\
 \ \ \ p_{W_1}p_{X_1^{k-1}|W_1}P_{X_{1,k}|W_1X_1^{k-1}}p_{S^{k-1}}p_{S_k}p_{S_{k+1}^n} \nonumber \\
\ \ \ \quad \cdot p_{W_2}p_{X_2^k|W_2X_1^{k-1}S^n}p_{Y^{k-1}|X_1^{k-1}X_2^{k-1}S^{k-1}}
\end{IEEEeqnarray*}
Summing this joint law over $w_1$ we obtain
\begin{IEEEeqnarray*}{l}
\sum_{w_1} p_{W_1W_2X_1^{k-1}X_{1,k}S^{k-1}S_kS_{k+1}^nX_2^kY^{k-1}} \nonumber \\
=  p_{W_2X_1^{k-1}X_{1,k}S^{k-1}S_kS_{k+1}^nX_2^kY^{k-1}} \nonumber \\
=p_{X_1^{k-1}}P_{X_{1,k}|X_1^{k-1}}p_{S^{k-1}}p_{S_k}p_{S_{k+1}^n} \nonumber \\
\ \ \ \quad \cdot p_{W_2}p_{X_2^k|W_2X_1^{k-1}S^n}p_{Y^{k-1}|X_1^{k-1}X_2^{k-1}S^{k-1}}
\end{IEEEeqnarray*}
Summing this joint law over all possible sub-sequences $(s_1,s_2,\ldots,s_{k-1})$ we obtain
\begin{IEEEeqnarray*}{l}
\sum_{(s_1,s_2,\ldots,s_{k-1})} p_{W_2X_1^{k-1}X_{1,k}S^{k-1}S_kS_{k+1}^nX_2^kY^{k-1}}
 \nonumber \\
=  p_{W_2X_1^{k-1}X_{1,k}S_kS_{k+1}^nX_2^kY^{k-1}} \nonumber \\
=p_{X_1^{k-1}}P_{X_{1,k}|X_1^{k-1}}p_{S_k}p_{S_{k+1}^n} \nonumber \\
\ \ \ \quad \cdot p_{W_2}p_{X_2^k|W_2X_1^{k-1}S_kS_{k+1}^n}p_{Y^{k-1}|X_1^{k-1}X_2^{k-1}}
\end{IEEEeqnarray*}
This establishes the Markov relation
\begin{IEEEeqnarray}{C}\label{eq:markov-Uk}
 X_{2,k}U_k\Markov S_kV_k\Markov X_{1,k}.
\end{IEEEeqnarray}

Next, let $J$ be a r.v. uniformly distributed over $\{1,\ldots,n\}$ and independent of
$(X_{1,k},X_{2,k},V_k,U_k,S_k,Y_k) \ , \ k=1,\ldots,n$, and define 
\begin{IEEEeqnarray*}{l}
(S,X_1,X_2,V,U,Y)=(S_J,X_{1,J},X_{2,J},V_J,U_J,Y_J).
\end{IEEEeqnarray*}
We may express (\ref{eq:R1upr1}) as follows
\begin{IEEEeqnarray}{l}
R_1\leq \frac{1}{n}\sum_{k=1}^n H(X_{1,k}|V_{k})=H(X_1|V,J)=H(X_1|\bar{V}),
\label{eq:R1upr11}
\end{IEEEeqnarray}
where in the last step we've defined $\bar{V}\triangleq (V,J)$. \\
Similarly, we may express (\ref{eq:R2upr1}) as follows
\begin{IEEEeqnarray}{rCl}
R_2 & \leq & \frac{1}{n} \sum_{k=1}^n \left[ I(U_k;Y_k|V_{k}X_{1,k})-I(U_k;S_k|V_k)\right] 
 \nonumber \\
& = & I(U;Y|V,X_1,J)-I(U;S|V,J) \nonumber \\
& = & I(U;Y|\bar{V},X_1)-I(U;S|\bar{V}) ,
\label{eq:R2upr11}
\end{IEEEeqnarray}
Finally, we may express (\ref{eq:sumRupr1}) as follows
\begin{IEEEeqnarray}{rCl}
R_1+R_2 & \leq & \frac{1}{n}  \sum_{k=1}^n \left[I(V_kU_kX_{1,k};Y_k)-I(U_k;S_k|V_k)\right] 
  \nonumber \\ 
& = & I(V,U,X_1;Y|J)-I(U;S|V,J) \nonumber \\
& = & I(V,J,U,X_1;Y)-I(J;Y)-I(U;S|V,J) \nonumber \\
& \leq & I(V,J,U,X_1;Y)-I(U;S|V,J) \nonumber \\
& = & I(\bar{V},U,X_1;Y)-I(U;S|\bar{V}).
\label{eq:sumRupr11}
\end{IEEEeqnarray}
This establishes the single letter expression for the achievable rate region (\ref{eq:r11}).
The convexity of the rate region (\ref{eq:r11}) can be shown in a similar way.

The inequalities (\ref{eq:R1upr1}), (\ref{eq:R2upr1}), 
(\ref{eq:sumRupr1}) combined with their respective single-letter expressions and the Markov relation  (\ref{eq:markov-Uk}) establish the converse part of Theorem~1.

\subsection{Bounds on alphabets sizes in Theorem~\ref{th:thm1}}
\label{sec:ba}

\vskip.1truein

We consider the alphabet sizes of $U$ and $V$. Specifically, let
$P_{X_1,X_2,S,V,U}$ be a distribution satisfying the Markov
conditions required in \eqref{eq:jointlawp}.
For convenience, $P_{X_1,X_2,S,U|V}(x_1,x_2,s,u|v)$ will be denoted
in the sequel as $P(\cdot|v)$.
We would like to bound the sizes of the alphabets ${\cal V}$ and ${\cal U}$,
while preserving the region given in \eqref{eq:r11}.
For a generic distribution $Q$ on ${\cal X}_1\times{\cal X}_2\times{\cal S}\times{\cal U}$,
define the functionals
\begin{subequations}
\label{eq:ba1}
\begin{IEEEeqnarray}{l}
q_{x_1,x_2,s}(Q) = \sum_u Q(x_1,x_2,s,u),\ \ \ x_1,x_2,s\in{\cal X}_1\times{\cal X}_2
 \times{\cal S}  \nonumber \\
  \label{eq:ba1_1}\\
J_1(Q) = \sum_{x_1,x_2,s,u} Q(x_1,x_2,s,u)\log \frac{1}{\sum_{x_2',s',u'}Q(x_1,x_2',s',u')}
 \nonumber \\ \label{eq:ba1_2}\\
J_2(Q) = \sum_{x_1,x_2,s,u} Q(x_1,x_2,s,u)\log \frac{1}{\sum_{x_1',x_2',u'}Q(x_1',x_2',s,u')}
 \nonumber \\ \label{eq:ba1_3}\\
J_3(Q) = \sum_{x_1,x_2,s,u} Q(x_1,x_2,s,u)\log \frac{\sum_{x_1',x_2',s'}Q(x_1',x_2',s',u)}
   {\sum_{x_1',x_2'}Q(x_1',x_2',s,u)}
 \nonumber \\ \label{eq:ba1_4}\\
J_4(Q) =  \sum_{x_1,x_2,s,u} Q(x_1,x_2,s,u)\nonumber\\
  \quad  \cdot\log \frac{1}{\sum_{x_2',s',u'}Q(x_1,x_2',s',u')   P_{Y|X_1,X_2,S}(y|x_1,x_2',s')}
 \nonumber \\ \label{eq:ba1_5}\\
J_5(Q) =  \sum_{x_1,x_2,s,u} Q(x_1,x_2,s,u)\nonumber\\
  \quad  \cdot\log \frac{1}{\sum_{x_2',s'}Q(x_1,x_2',s',u)   P_{Y|X_1,X_2,S}(y|x_1,x_2',s')}
 \nonumber \\ \label{eq:ba1_6}  .
\end{IEEEeqnarray}
\end{subequations}
Substituting the distribution $P_{X_1,X_2,S,U|V}(\cdot|v)$ in the functionals,
and averaging them with respect to $v$, we obtain
\begin{subequations}
\label{eq:ba2}
\begin{IEEEeqnarray}{rCl}
\sum_v P_V(v)q_{x_1,x_2,s}(P(\cdot|v)) &=& P_{X_1,X_2,S}(x_1,x_2,s)
\label{eq:ba2_1}\\
\sum_v P_V(v)J_1(P(\cdot|v)) &=& H(X_1|V)\label{eq:ba2_2}\\
\sum_v P_V(v)J_2(P(\cdot|v)) &=& H(S|V)\label{eq:ba2_3}\\
\sum_v P_V(v)J_3(P(\cdot|v)) &=& H(S|U,V)\label{eq:ba2_4}\\
\sum_v P_V(v)J_4(P(\cdot|v)) &=& H(Y|V,X_1)\label{eq:ba2_5}\\
\sum_v P_V(v)J_5(P(\cdot|v)) &=& H(Y|V,U,X_1)\label{eq:ba2_6}.
\end{IEEEeqnarray}
\end{subequations}
Observe that preserving the values of the right hand sides of
\eqref{eq:ba2_1}-\eqref{eq:ba2_6}, guarantees that we also preserve the
region \eqref{eq:r11}. We used here the Markov
structure $Y \Markov (X_1,X_2,S) \Markov (V,U)$, and the fact that if we preserve the joint
distribution of $X_1,X_2,S$, the distribution of $Y$  is
also preserved. By the Support Lemma~\cite{CKBook},
we can restrict the alphabet of $V$ to:
\be
|{\cal V}| \leq |{\cal X}_1| |{\cal X}_2| |{\cal S}|  + 5. \label{eq:ba3}
\ee
Note that this bound is independent of the alphabet of $U$. 

We now fix some $V$ with bounded alphabet as above, and proceed to bound the
alphabet of $U$. Let $\tilde{Q}$ be a generic distribution on
${\cal X}_1\times{\cal X}_2\times{\cal S}\times{\cal V}$, and define the functionals
\begin{subequations}
\label{eq:ba4}
\begin{IEEEeqnarray}{l}
\tilde{q}_{x_1,x_2,s,v}(\tilde{Q}) = \tilde{Q}(x_1,x_2,s,v)\label{eq:ba4_1}\\
\tilde{J}_1(\tilde{Q}) = \sum_{x_1,x_2,s,v}\tilde{Q}(x_1,x_2,s,v)\nonumber\\
 \ \ \ \ \quad   \cdot\log\frac{\sum_{x_1',x_2',s'}\tilde{Q}(x_1',x_2',s',v)}
    {\sum_{x_1',x_2'}\tilde{Q}(x_1',x_2',s,v)}
   \nonumber \\ \label{eq:ba4_2}\\
\tilde{J}_2(\tilde{Q}) = \sum_{x_1,x_2,s,v}\tilde{Q}(x_1,x_2,s,v)\nonumber\\
 \quad   \cdot\log\frac{1}{\sum_{x_2',s'}\tilde{Q}(x_1,x_2',s',v)P_{Y|X_1,X_2,S}(y|x_1,x_2',s')}
   \nonumber \\ \label{eq:ba4_3} .
\end{IEEEeqnarray}
\end{subequations}
Since
\begin{IEEEeqnarray*}{rCl}
P_{Y|V,X_1}(y|v,x_1) &=& \sum_{x_2,s}P_{Y|X_1,X_2,S}(y|x_1,x_2,s)\\
  & & \cdot\frac{P_{X_1,X_2,S,V}(x_1,x_2,s,v)}
   {\sum_{x_2',s'}P_{X_1,X_2,S,V}(x_1,x_2',s',v)},
\end{IEEEeqnarray*}
in order to preserve the value of $H(Y|V,X_1)$, it suffices to
preserve the joint distribution of $X_1,X_2,S,V$.

For convenience, we use in  the sequel the shorthand notation
$P(\cdot|u)=P_{X_1,X_2,S,V|U}(\cdot|u)$.
Substituting the distribution
$P_{X_1,X_2,S,V|U}(\cdot|u)$ in the functionals~\eqref{eq:ba4}
and averaging over $u$, we obtain
\begin{subequations}
\label{eq:ba5}
\begin{IEEEeqnarray}{rCl}
\sum_u P_U(u)\tilde{q}_{x_1,x_2,s,v}(P(\cdot|u)) &=& P_{X_1,X_2,S,V}(x_1,x_2,s,v)
 \nonumber \\
  \label{eq:ba5_1}\\
\sum_u P_U(u)\tilde{J}_1(P(\cdot|u)) &=& H(S|V,U)
                              \label{eq:ba5_2}\\
\sum_u P_U(u)\tilde{J}_2(P(\cdot|u)) &=& H(Y|V,U,X_1)
                              \label{eq:ba5_3}
\end{IEEEeqnarray}
\end{subequations}
Applying again the Support Lemma, we see that it suffices to bound the
alphabe size of $U$ as
\be
|{\cal U}|\leq |{\cal X}_1| |{\cal X}_2| |{\cal S}|  |{\cal V}| + 2.
\ee
This completes the proof of the bounds on the alphabet sizes.

\subsection{Proof of Theorem~\ref{th:thm2}}

The achievability part follows similarly to that of Theorem~\ref{th:thm1} the only difference being in the way the codeword 
$\mbox{\boldmath $x$}_2({\mbox{\boldmath $s$}}^{(b)},w_{2}^{(b)},\omega_0^{(b)})$ is generated. Here the second encoder generates the codeword 
$\mbox{\boldmath $x$}_2({\mbox{\boldmath $s$}}^{(b)},w_{2}^{(b)},\omega_0^{(b)})$ by
drawing its components i.i.d. conditionally on the quadruple $({\mbox{\boldmath $s$}}^{(b)},{\mbox{\boldmath $u$}}(w_{2}^{(b)},\jmath_0,\omega_0^{(b)}),{\mbox{\boldmath $v$}}(\omega_0^{(b)}),
 {\mbox{\boldmath $x$}}_1^{(b)})$,
where the conditional law is induced by (\ref{eq:jointlawpc}).  

\medskip

For the converse, consider an $(e^{nR_1},e^{nR_2},n)$ code with average block error
probability $P_e^{(n)}$, and a law on ${\cal W}_1\times{\cal
W}_2\times{\cal X}_1^n\times{\cal X}_2^n\times{\cal Y}^n\times{\cal
S}^n$ given by
\begin{IEEEeqnarray}{l}
p_{W_1W_2X_1^nX_2^nS^nY^n} \nonumber \\
=p_{W_1}p_{W_2}I_{\{X_1^n=f_1(W_1)\}}p_{X_2^n|W_1W_2S^n}
\prod_{k=1}^np_{Y_k|X_{1,k}X_{2,k}S_k}. \nonumber \\
\label{eq:convlawc}
\end{IEEEeqnarray}

\medskip

The Fano inequalities for the causal cribbing case yield the same inequalities (\ref{eq:R1upr1}), (\ref{eq:R2upr1}),  and (\ref{eq:sumRupr1}). 

\medskip

It remains to verify the joint law of the auxiliary random variables.

By \eqref{eq:convlawc} and the encoding rule (\ref{eq:enc2c}) we may write
\begin{IEEEeqnarray*}{l}
p_{W_1W_2X_1^{k-1}X_{1,k}S^{k-1}S_kS_{k+1}^nX_2^kY^{k-1}}= \nonumber \\
 \ \ \ p_{W_1}p_{X_1^{k-1}|W_1}P_{X_{1,k}|W_1X_1^{k-1}}p_{S^{k-1}}p_{S_k}p_{S_{k+1}^n} \nonumber \\
\ \ \ \quad \cdot p_{W_2}p_{X_2^{k-1}|W_2X_1^{k-1}S^n}
  p_{X_{2,k}|W_2X_1^{k-1}X_{1,k}X_2^{k-1}S^n}  \nonumber \\
\ \ \ \quad \cdot p_{Y^{k-1}|X_1^{k-1}X_2^{k-1}S^{k-1}}
\end{IEEEeqnarray*}
Summing this joint law over $w_1$ we obtain
\begin{IEEEeqnarray*}{l}
\sum_{w_1} p_{W_1W_2X_1^{k-1}X_{1,k}S^{k-1}S_kS_{k+1}^nX_2^kY^{k-1}} \nonumber \\
=  p_{W_2X_1^{k-1}X_{1,k}S^{k-1}S_kS_{k+1}^nX_2^kY^{k-1}} \nonumber \\
=p_{X_1^{k-1}}P_{X_{1,k}|X_1^{k-1}}p_{S^{k-1}}p_{S_k}p_{S_{k+1}^n} \nonumber \\
\ \ \ \quad \cdot p_{W_2}p_{X_2^{k-1}|W_2X_1^{k-1}S^n}
  p_{X_{2,k}|W_2X_1^{k-1}X_{1,k}X_2^{k-1}S^n}  \nonumber \\
\ \ \ \quad \cdot p_{Y^{k-1}|X_1^{k-1}X_2^{k-1}S^{k-1}}
\end{IEEEeqnarray*}
Summing this joint law over all possible sub-sequences $(s_1,s_2,\ldots,s_{k-1})$ we obtain
\begin{IEEEeqnarray*}{l}
\sum_{(s_1,s_2,\ldots,s_{k-1})} p_{W_2X_1^{k-1}X_{1,k}S^{k-1}S_kS_{k+1}^nX_2^kY^{k-1}}
 \nonumber \\
=  p_{W_2X_1^{k-1}X_{1,k}S_kS_{k+1}^nX_2^kY^{k-1}} \nonumber \\
=p_{X_1^{k-1}}P_{X_{1,k}|X_1^{k-1}}p_{S_k}p_{S_{k+1}^n} \nonumber \\
\ \ \quad \cdot p_{W_2}p_{X_2^{k-1}|W_2X_1^{k-1}S_kS_{k+1}^n}
  p_{X_{2,k}|W_2X_1^{k-1}X_{1,k}X_2^{k-1}S_kS_{k+1}^n}  \nonumber \\
\ \ \ \quad \cdot p_{Y^{k-1}|X_1^{k-1}X_2^{k-1}}
\end{IEEEeqnarray*}
This establishes the Markov relation
\begin{IEEEeqnarray}{C}\label{eq:markov-Ukc}
 U_k\Markov S_kV_k\Markov X_{1,k} ,
\end{IEEEeqnarray}
as well as the fact that conditionally on $V_kU_kS_kX_{1,k}$ the r.v. $X_{2,k}$ 
is independent of the rest. 

\begin{appendix}
\subsection{Strong Typicality}\label{sec:st}
Let $\left\{X^{(1)},X^{(2)},\ldots,X^{(k)}\right\}$ denote a finite
collection of discrete random variables with some joint distribution
$P\left(x^{(1)},x^{(2)},\ldots,x^{(k)}\right)$ with
$\left(x^{(1)},x^{(2)},\ldots,x^{(k)}\right)\in{\cal
X}^{(1)}\times{\cal X}^{(2)}\times\ldots\times{\cal X}^{(k)}$. Let
$S$ denote an ordered nonempty subset of these random variables and
consider $n$ independent copies of $S$. Thus, with $\mbox{\boldmath
$S$}\triangleq(S_1,S_2,\ldots,S_n)$,
\begin{eqnarray*}
\Pr\{\mbox{\boldmath $S$}=\mbox{\boldmath $s$}\}=\prod_{j=1}^n
\Pr\{S_j=s_j\}.
\end{eqnarray*}
Let $N(s;\mbox{\boldmath $s$})$ be the number of indices
$j\in\{1,2,\ldots,n\}$ such that $S_{j}=s$.  By the law of large
numbers, for any subset $S$ of random variables and for all $s\in
S$,
\begin{eqnarray}
\frac{1}{n}N(s;\mbox{\boldmath $s$})\rightarrow P(s),
\label{eq:conv1}
\end{eqnarray}
as well as
\begin{eqnarray}
-\frac{1}{n}\ln
P(s_{1},s_{2},\ldots,s_{n})=-\frac{1}{n}\sum_{j=1}^{n}\ln
P(s_{j})\rightarrow H(S). \label{eq:conv2}
\end{eqnarray}
The convergence in (\ref{eq:conv1}) and (\ref{eq:conv2}) takes place
simultaneously with probability one for all nonempty subsets $S$
\cite{cov}.

\begin{definition}The set ${\cal A}_{\epsilon}$ of
$\epsilon$-strongly typical $n$-sequences is defined by (see
\cite[Chapter 3,12,13]{cov})
\begin{eqnarray*}
{\cal A}_{\epsilon} & \triangleq & {\cal
A}_{\epsilon}\left(X^{(1)},X^{(2)},\ldots,X^{(k)}\right)
  \\ &\triangleq  & \Bigg\{\left(\mbox{\boldmath $x$}^{(1)},\mbox{\boldmath
$x$}^{(2)},\ldots,\mbox{\boldmath $x$}^{(k)}\right): \\
& & \quad  \Big|\frac{1}{n}N\left(x^{(1)},x^{(2)},\ldots,x^{(k)};
   \mbox{\boldmath $x$}^{(1)},\mbox{\boldmath $x$}^{(2)},\ldots,\mbox{\boldmath
$x$}^{(k)}\right) \\
& &\quad \qquad  -
   P\left(x^{(1)},x^{(2)},\ldots,x^{(k)}\right)\Big| \\
& & \quad <\frac{\epsilon}{\|
   {\cal X}^{(1)}\times{\cal X}^{(2)}\times\ldots\times{\cal X}^{(k)}\|}, \\
& & \quad  \forall\left(\mbox{\boldmath $x$}^{(1)},\mbox{\boldmath
$x$}^{(2)},\ldots,\mbox{\boldmath $x$}^{(k)}\right)\in {\cal
X}^{(1)}\times 
\ldots\times{\cal X}^{(k)} \Bigg\},
\end{eqnarray*}
where $\|{\cal X}\|$ is the cardinality of the set ${\cal X}$.\\
Let ${\cal A}_{\epsilon}(S)$ be defined similar to ${\cal
A}_{\epsilon}$ but now with constraints corresponding to all
nonempty subsets of $S$. We recall now two basic lemmas (for the
proofs we refer to \cite{cov}).
\end{definition}

\begin{lemma} \label{th:lm1}For any $\epsilon>0$ the following
statements hold for every integer $n \geq 1$:
\begin{enumerate}
\item If $\mbox{\boldmath $s$}\in{\cal A}_{\epsilon}(S)$, then
$\exp\left(-n(H(S)+\epsilon)\right)\leq
  \Pr\{\mbox{\boldmath $S$}=\mbox{\boldmath $s$}\}\leq
\exp\left(-n(H(S)-\epsilon)\right)$.
\item If $S_1,S_2\subseteq \left\{X_1,X_2,\ldots,X_k\right\}$ and
  $(\mbox{\boldmath $s$}_1,\mbox{\boldmath $s$}_2)\in{\cal
A}_{\epsilon}(S_1\cup S_2)$, then
  \begin{eqnarray*}\lefteqn{
  \exp\left(-n(H(S_1|S_2)+2\epsilon)\right)\leq
  \Pr\{\mbox{\boldmath $S_1$}=\mbox{\boldmath $s_1$}|\mbox{\boldmath
$S_2$}=\mbox{\boldmath $s_2$}\}} \\
& &
  \leq \exp\left(-n(H(S_1|S_2)-2\epsilon)\right) .\hspace{2cm}
  \end{eqnarray*}
Moreover, the following statements hold for every sufficiently large
  $n$:
\item $\Pr\left\{{\cal A}_{\epsilon}(S)\right\}\geq 1-\epsilon$,
\item $(1-\epsilon)\exp(n(H(S)-\epsilon))\leq \left\|{\cal
A}_{\epsilon}(S)\right\|\leq \exp(n(H(S)+\epsilon)).$
\end{enumerate}

\end{lemma}
\vskip.1truein
\begin{lemma}\label{th:lm2} Let the discrete random variables
$X,Y,Z$ have joint distribution $P_{X,Y,Z}(x,y,z)$. Let $X'$ and
$Y'$ be conditionally independent given $Z$, with the marginal laws
\begin{eqnarray*}
P_{X'|Z}(x|z) & = & \sum_{y} P_{X,Y,Z}(x,y,z)/P_{Z}(z)  , \\
P_{Y'|Z}(y|z) & = & \sum_{x}P_{X,Y,Z}(x,y,z)/P_{Z}(z)  .
\end{eqnarray*}
Let $(\mbox{\boldmath $X$},\mbox{\boldmath $Y$},\mbox{\boldmath
  $Z$})\sim\prod_{k=1}^n P_{X,Y,Z}(x_k,y_k,z_k)$ and
$(\mbox{\boldmath $X$}',\mbox{\boldmath $Y$}', \mbox{\boldmath
$Z$})\sim\prod_{k=1}^n P_{X'|Z}(x_k|z_k)
  P_{Y'|Z}(y_k|z_k) P_{Z}(z_k)$.
Then
\begin{equation*}
\Pr\left\{(\mbox{\boldmath $X$}',\mbox{\boldmath
$Y$}',\mbox{\boldmath $Z$})\in{\cal A}_{\epsilon}(X,Y,Z)\right\}
\leq\exp(-n[I(X;Y|Z)-\epsilon]).
\end{equation*}
 \end{lemma}

\end{appendix}


\begin{thebibliography}{99}
\bibitem{anelia} A. Somekh-Baruch, S. Shamai (Shitz)
 and S. Verd\'{u}, ``Cooperative multiple-access encoding with states available
at one transmitter,'' {\it IEEE Trans. Inform. Theory,} vol. IT-54, no. 10, pp.
4448-4469, Oct. 2008.


\bibitem{frans} F.M.J. Willems and E.C. van der Meulen, ``The discrete
  memoryless multiple-access channel with cribbing encoders'', {\it
  IEEE Trans. Inform. Theory,} vol. IT-31, no. 3, pp. 313-327, May 1985.

\bibitem{rimoldiurbanke96} B. Rimoldi and R. Urbanke, ``A
  rate-splitting approach to the {G}aussian multiple-access channel'',
  {\it IEEE Trans. Inform. Theory,} vol. IT-42, no. 2, pp. 364 -- 375,
  Mar 1996.

\bibitem{wozencraftjacobs65} J. M. Wozencraft and I. M. Jacobs, {\it
    Principles of Communication Engineering.} John Wiley \& Sons,
    1965.
\bibitem{csi} I. Csisz\'{a}r and J. K\"{o}rner, ``Broadcast channels with
 confidential messages,''
 {\it IEEE Trans. Inform. Theory,} vol. IT-24, No. 3 pp. 339-348, May 1978.


\bibitem{ahlkor} R. Ahlswede and J. K\"{o}rner, ``Source coding with side
information and a converse for degraded broadcast channels,'' {\it
IEEE Trans. Inform. Theory,} vol. IT-21, No. 6, pp. 629-637, Nov.
1975.
\bibitem{wy1} A. D. Wyner, ``On source coding with side information at the
decoder,'' {\it IEEE Trans. Inform. Theory,} vol. IT-21, No. 6,
pp. 294-300, May 1975.
\bibitem{ber} T. Berger, ``Multiterminal source coding.'' Lecture notes
presented at the 1977 CISM Summer School, Udine, Italy, July
18-20, 1977, Springer-Verlag.


\bibitem{cov} T. M. Cover and J. A. Thomas, ``{\it Elements of Information
 Theory,}'' Wiley, 1991.

\bibitem{CKBook}  I. Csisz\'{a}r and J. K\"{o}rner,
 \emph{Information Theory: Coding Theorems for Discrete Memoryless Systems.} New York: Academic, 1981.

\end{thebibliography}
\end{document}